
\font\twelverm=cmr10 scaled 1200    \font\twelvei=cmmi10 scaled 1200
\font\twelvesy=cmsy10 scaled 1200   \font\twelveex=cmex10 scaled 1200
\font\twelvebf=cmbx10 scaled 1200   \font\twelvesl=cmsl10 scaled 1200
\font\twelvett=cmtt10 scaled 1200   \font\twelveit=cmti10 scaled 1200

\skewchar\twelvei='177   \skewchar\twelvesy='60


\def\twelvepoint{\normalbaselineskip=12.4pt
  \abovedisplayskip 12.4pt plus 3pt minus 9pt
  \belowdisplayskip 12.4pt plus 3pt minus 9pt
  \abovedisplayshortskip 0pt plus 3pt
  \belowdisplayshortskip 7.2pt plus 3pt minus 4pt
  \smallskipamount=3.6pt plus1.2pt minus1.2pt
  \medskipamount=7.2pt plus2.4pt minus2.4pt
  \bigskipamount=14.4pt plus4.8pt minus4.8pt
  \def\rm{\fam0\twelverm}          \def\it{\fam\itfam\twelveit}%
  \def\sl{\fam\slfam\twelvesl}     \def\bf{\fam\bffam\twelvebf}%
  \def\mit{\fam 1}                 \def\cal{\fam 2}%
  \def\tt{\twelvett}
  \def\nullspace{\nulldelimiterspace=0pt \mathsurround=0pt }
  \def\big##1{{\hbox{$\left##1\vbox to 10.2pt{}\right.\nullspace$}}}
  \def\Big##1{{\hbox{$\left##1\vbox to 13.8pt{}\right.\nullspace$}}}
  \def\bigg##1{{\hbox{$\left##1\vbox to 17.4pt{}\right.\nullspace$}}}
  \def\Bigg##1{{\hbox{$\left##1\vbox to 21.0pt{}\right.\nullspace$}}}
  \textfont0=\twelverm   \scriptfont0=\tenrm   \scriptscriptfont0=\sevenrm
  \textfont1=\twelvei    \scriptfont1=\teni    \scriptscriptfont1=\seveni
  \textfont2=\twelvesy   \scriptfont2=\tensy   \scriptscriptfont2=\sevensy
  \textfont3=\twelveex   \scriptfont3=\twelveex  \scriptscriptfont3=\twelveex
  \textfont\itfam=\twelveit
  \textfont\slfam=\twelvesl
  \textfont\bffam=\twelvebf
  \scriptfont\bffam=\tenbf
  \scriptscriptfont\bffam=\sevenbf
  \normalbaselines\rm}



\def\beginlinemode{\endmode
  \begingroup\parskip=0pt \obeylines\def\\{\par}\def\endmode{\par\endgroup}}
\def\beginparmode{\endmode
  \begingroup \def\endmode{\par\endgroup}}
\let\endmode=\par
{\obeylines\gdef\
{}}
\def\singlespace{\baselineskip=\normalbaselineskip}
\def\oneandahalfspace{\baselineskip=\normalbaselineskip
  \multiply\baselineskip by 3 \divide\baselineskip by 2}
\def\oneandathirdspace{\baselineskip=\normalbaselineskip
  \multiply\baselineskip by 4 \divide\baselineskip by 3}
\def\doublespace{\baselineskip=\normalbaselineskip \multiply\baselineskip by 2}
\newcount\firstpageno
\firstpageno=2
\footline={\ifnum\pageno<\firstpageno{\hfil}\else{\hfil\twelverm\folio\hfil}\fi}
\let\rawfootnote=\footnote
\def\footnote#1#2{{\tenrm\singlespace\parindent=0pt\rawfootnote{#1}{#2}}}
\def\raggedcenter{\leftskip=4em plus 12em \rightskip=\leftskip
  \parindent=0pt \parfillskip=0pt \spaceskip=.3333em \xspaceskip=.5em
  \pretolerance=9999 \tolerance=9999
  \hyphenpenalty=9999 \exhyphenpenalty=9999 }
\def\dateline{\rightline{\ifcase\month\or
  January\or February\or March\or April\or May\or June\or
  July\or August\or September\or October\or November\or December\fi
  \space\number\year}}
\def\received{\vskip 3pt plus 0.2fill
 \centerline{\sl (Received\space\ifcase\month\or
  January\or February\or March\or April\or May\or June\or
  July\or August\or September\or October\or November\or December\fi
  \qquad, \number\year)}}


\hsize=16truecm 
\hoffset=0.0truecm
\vsize=25truecm
\voffset=0.0truecm
\parskip=0pt
\twelvepoint  
\oneandathirdspace
\overfullrule=0pt 



\def\title
  {\null\vskip 3pt plus 0.2fill
   \beginlinemode \doublespace \raggedcenter \bf}

\font\titlefont=cmr10 scaled \magstep3


\font\twelvesc=cmcsc10 scaled 1200
\def\author{\vskip 16pt plus 0.2fill \beginlinemode\singlespace
\raggedcenter\twelvesc}

\def\affil
  {\vskip 4pt plus 0.1fill \beginlinemode
   \oneandahalfspace \raggedcenter \sl}

\def\abstract  
  {\vskip 24pt plus 0.3fill \beginparmode
   \narrower\centerline{ABSTRACT}\vskip 12pt }

\def\endpage{\vfill\eject}     

\def\body{\beginparmode}

\def\endtitlepage{\endpage\body}

\def\head#1{
  \filbreak\vskip 0.35truein
  {\immediate\write16{#1}
   \raggedcenter \uppercase{#1}\par}
   \nobreak\vskip 0.2truein\nobreak}

\def\subhead#1{
  \vskip 0.20truein
  {\raggedcenter #1 \par}
   \nobreak\vskip 0.15truein\nobreak}

\def\References  
  {
   \subhead{References}
   \beginparmode
   \frenchspacing\parindent=0pt \leftskip=1truecm
   \parskip=2pt plus 3pt \everypar{\hangindent=\parindent}}

\gdef\refis#1{\indent\hbox to 0pt{\hss#1.~}}   

\def\endreferences{\body}


\def\figurecaptions
  {\endpage
   \beginparmode
   \head{Figure Captions}
}



\def\endpaper{\endmode\vfill\supereject}


\def\cite#1{{#1}}
\def\[#1]{[\cite{#1}]}       
\def\refto#1{$^{{#1}}$}      
\def\ref#1{Ref.~#1}          
\def\Ref#1{Ref.~#1}          

\def\call#1{{#1}}
\def\(#1){(\call{#1})}
\def\Eq#1{eq.~\(#1)}                   
\def\Eqs#1{eqs.~\(#1)}                 

\catcode`@=11
\newcount\tagnumber\tagnumber=0

\immediate\newwrite\eqnfile
\newif\if@qnfile\@qnfilefalse
\def\write@qn#1{}
\def\writenew@qn#1{}
\def\w@rnwrite#1{\write@qn{#1}\message{#1}}
\def\@rrwrite#1{\write@qn{#1}\errmessage{#1}}

\def\t@ghead{}
\def\taghead#1{\gdef\t@ghead{#1}\global\tagnumber=0}

\expandafter\def\csname @qnnum-3\endcsname
  {{\t@ghead\advance\tagnumber by -3\relax\number\tagnumber}}
\expandafter\def\csname @qnnum-2\endcsname
  {{\t@ghead\advance\tagnumber by -2\relax\number\tagnumber}}
\expandafter\def\csname @qnnum-1\endcsname
  {{\t@ghead\advance\tagnumber by -1\relax\number\tagnumber}}
\expandafter\def\csname @qnnum0\endcsname
  {\t@ghead\number\tagnumber}
\expandafter\def\csname @qnnum+1\endcsname
  {{\t@ghead\advance\tagnumber by 1\relax\number\tagnumber}}
\expandafter\def\csname @qnnum+2\endcsname
  {{\t@ghead\advance\tagnumber by 2\relax\number\tagnumber}}
\expandafter\def\csname @qnnum+3\endcsname
  {{\t@ghead\advance\tagnumber by 3\relax\number\tagnumber}}

\def\equationfile{%
  \@qnfiletrue\immediate\openout\eqnfile=\jobname.eqn%
  \def\write@qn##1{\if@qnfile\immediate\write\eqnfile{##1}\fi}
  \def\writenew@qn##1{\if@qnfile\immediate\write\eqnfile
    {\noexpand\tag{##1} = (\t@ghead\number\tagnumber)}\fi}
}

\def\callall#1{\xdef#1##1{#1{\noexpand\call{##1}}}}
\def\call#1{\each@rg\callr@nge{#1}}

\def\each@rg#1#2{{\let\thecsname=#1\expandafter\first@rg#2,\end,}}
\def\first@rg#1,{\thecsname{#1}\apply@rg}
\def\apply@rg#1,{\ifx\end#1\let\next=\relax%
\else,\thecsname{#1}\let\next=\apply@rg\fi\next}

\def\callr@nge#1{\calldor@nge#1-\end-}
\def\callr@ngeat#1\end-{#1}
\def\calldor@nge#1-#2-{\ifx\end#2\@qneatspace#1 %
  \else\calll@@p{#1}{#2}\callr@ngeat\fi}
\def\calll@@p#1#2{\ifnum#1>#2{\@rrwrite{Equation range #1-#2\space is bad.}
\errhelp{If you call a series of equations by the notation M-N, then M and
N must be integers, and N must be greater than or equal to M.}}\else%
 {\count0=#1\count1=#2\advance\count1
by1\relax\expandafter\@qncall\the\count0,%
  \loop\advance\count0 by1\relax%
    \ifnum\count0<\count1,\expandafter\@qncall\the\count0,%
  \repeat}\fi}

\def\@qneatspace#1#2 {\@qncall#1#2,}
\def\@qncall#1,{\ifunc@lled{#1}{\def\next{#1}\ifx\next\empty\else
  \w@rnwrite{Equation number \noexpand\(>>#1<<) has not been defined yet.}
  >>#1<<\fi}\else\csname @qnnum#1\endcsname\fi}

\let\eqnono=\eqno
\def\eqno(#1){\tag#1}
\def\tag#1$${\eqnono(\displayt@g#1 )$$}

\def\aligntag#1\endaligntag
  $${\gdef\tag##1\\{&(##1 )\cr}\eqalignno{#1\\}$$
  \gdef\tag##1$${\eqnono(\displayt@g##1 )$$}}

\def\eqalignno#1{\displ@y \tabskip\centering
  \halign to\displaywidth{\hfil$\displaystyle{##}$\tabskip\z@skip
    &$\displaystyle{{}##}$\hfil\tabskip\centering
    &\llap{$\displayt@gpar##$}\tabskip\z@skip\crcr
    #1\crcr}}

\def\displayt@gpar(#1){(\displayt@g#1 )}

\def\displayt@g#1 {\rm\ifunc@lled{#1}\global\advance\tagnumber by1
        {\def\next{#1}\ifx\next\empty\else\expandafter
        \xdef\csname @qnnum#1\endcsname{\t@ghead\number\tagnumber}\fi}%
  \writenew@qn{#1}\t@ghead\number\tagnumber\else
        {\edef\next{\t@ghead\number\tagnumber}%
        \expandafter\ifx\csname @qnnum#1\endcsname\next\else
        \w@rnwrite{Equation \noexpand\tag{#1} is a duplicate number.}\fi}%
  \csname @qnnum#1\endcsname\fi}

\def\ifunc@lled#1{\expandafter\ifx\csname @qnnum#1\endcsname\relax}

\let\@qnend=\end\gdef\end{\if@qnfile
\immediate\write16{Equation numbers written on []\jobname.EQN.}\fi\@qnend}

\catcode`@=12


\catcode`@=11
\newcount\r@fcount \r@fcount=0
\newcount\r@fcurr
\immediate\newwrite\reffile
\newif\ifr@ffile\r@ffilefalse
\def\w@rnwrite#1{\ifr@ffile\immediate\write\reffile{#1}\fi\message{#1}}

\def\writer@f#1>>{}
\def\referencefile{
  \r@ffiletrue\immediate\openout\reffile=\jobname.ref%
  \def\writer@f##1>>{\ifr@ffile\immediate\write\reffile%
    {\noexpand\refis{##1} = \csname r@fnum##1\endcsname = %
     \expandafter\expandafter\expandafter\strip@t\expandafter%
     \meaning\csname r@ftext\csname r@fnum##1\endcsname\endcsname}\fi}%
  \def\strip@t##1>>{}}

\def\citeall#1{\xdef#1##1{#1{\noexpand\cite{##1}}}}
\def\cite#1{\each@rg\citer@nge{#1}}

\def\each@rg#1#2{{\let\thecsname=#1\expandafter\first@rg#2,\end,}}
\def\first@rg#1,{\thecsname{#1}\apply@rg}
\def\apply@rg#1,{\ifx\end#1\let\next=\relax
\else,\thecsname{#1}\let\next=\apply@rg\fi\next}

\def\citer@nge#1{\citedor@nge#1-\end-}
\def\citer@ngeat#1\end-{#1}
\def\citedor@nge#1-#2-{\ifx\end#2\r@featspace#1 
  \else\citel@@p{#1}{#2}\citer@ngeat\fi}
\def\citel@@p#1#2{\ifnum#1>#2{\errmessage{Reference range #1-#2\space is bad.}%
    \errhelp{If you cite a series of references by the notation M-N, then M and
    N must be integers, and N must be greater than or equal to M.}}\else%
 {\count0=#1\count1=#2\advance\count1
by1\relax\expandafter\r@fcite\the\count0,%
  \loop\advance\count0 by1\relax
    \ifnum\count0<\count1,\expandafter\r@fcite\the\count0,%
  \repeat}\fi}

\def\r@featspace#1#2 {\r@fcite#1#2,}
\def\r@fcite#1,{\ifuncit@d{#1}
    \newr@f{#1}%
    \expandafter\gdef\csname r@ftext\number\r@fcount\endcsname%
                     {\message{Reference #1 to be supplied.}%
                      \writer@f#1>>#1 to be supplied.\par}%
 \fi%
 \csname r@fnum#1\endcsname}
\def\ifuncit@d#1{\expandafter\ifx\csname r@fnum#1\endcsname\relax}%
\def\newr@f#1{\global\advance\r@fcount by1%
    \expandafter\xdef\csname r@fnum#1\endcsname{\number\r@fcount}}

\let\r@fis=\refis
\def\refis#1#2#3\par{\ifuncit@d{#1}
   \newr@f{#1}%
   \w@rnwrite{Reference #1=\number\r@fcount\space is not cited up to now.}\fi%
  \expandafter\gdef\csname r@ftext\csname r@fnum#1\endcsname\endcsname%
  {\writer@f#1>>#2#3\par}}

\def\ignoreuncited{
   \def\refis##1##2##3\par{\ifuncit@d{##1}%
     \else\expandafter\gdef\csname r@ftext\csname
r@fnum##1\endcsname\endcsname%

     {\writer@f##1>>##2##3\par}\fi}}

\def\r@ferr{\endreferences\errmessage{I was expecting to see
\noexpand\endreferences before now;  I have inserted it here.}}
\let\r@ferences=\references
\def\references{\r@ferences\def\endmode{\r@ferr\par\endgroup}}

\let\endr@ferences=\endreferences
\def\endreferences{\r@fcurr=0
  {\loop\ifnum\r@fcurr<\r@fcount
   \advance\r@fcurr by 1\relax\expandafter\r@fis\expandafter{\number\r@fcurr}%
    \csname r@ftext\number\r@fcurr\endcsname%
  \repeat}\gdef\r@ferr{}\endr@ferences}


\let\r@fend=\endpaper\gdef\endpaper{\ifr@ffile
\immediate\write16{Cross References written on []\jobname.REF.}\fi\r@fend}

\catcode`@=12

\citeall\refto
\citeall\ref%
\citeall\Ref%

\null
{\oneandahalfspace
\centerline {\titlefont Gravitational Lenses and Unconventional Gravity
Theories} \vskip 36pt
\centerline{Jacob D. Bekenstein}
\vskip 15pt
\centerline{\it The Racah Institute of Physics, Hebrew University
of Jerusalem,\/} \centerline{{\it Givat Ram, Jerusalem 91904,
Israel\/}\footnote{$^1$}{E-mail: bekenste@vms.huji.ac.il\hfill}}
\vskip 20 pt     \centerline{{\twelverm and}} \vskip 20 pt
\centerline{Robert H. Sanders}
\vskip 15pt
\centerline{\it Kapteyn Astronomical Institute, University
of Groningen,\/} \centerline{{\it 9700 AV Groningen, The
Netherlands}\footnote{$^2$}{E-mail: sanders@astro.rug.nl\hfill}}}
\vskip 36pt

We study gravitational lensing by clusters of galaxies in the context of the
generic class of unconventional gravity theories which describe gravity in
terms of a metric and one or more  scalar fields (called here scalar--tensor
theories).  We conclude that if the scalar fields have positive energy, then
whatever their dynamics,  the bending of light by a weakly gravitating system,
like a galaxy or a cluster of galaxies, cannot exceed the bending predicted by
general relativity for the mass of visible and hitherto undetected matter (but
excluding the scalar field's energy).  Thus use of general relativity to
interpret gravitational lensing observations can only underestimate the mass
present in stars, gas and dark matter. The same conclusion obtains within
general relativity if a nonnegligible part of the mass in clusters is in the
form of coherent scalar fields, {\it i.e.\/} Higgs fields.  The popular
observational claim that clusters of galaxies deflect light much more strongly
than would be expected from the observable  matter contained by them, if it
survives, cannot be interpreted in terms of some scalar-tensor unconventional
gravity theory with no dark matter.  And  if the observations eventually show
that the matter distribution inferred via general relativity from the lensing
is very much like that determined from the dynamics of test objects, then
scalar--tensor unconventional gravity will be irrelevant for understanding the
mass discrepancy in clusters.  However, even a single system in which the
dynamical mass determined from virial methods significantly exceeds the
lensing mass as determined by general relativity, would be very problematic
for the dark matter picture, but would be entirely consistent with
unconventional scalar--tensor gravity theory.

\endtitlepage

\centerline{1. INTRODUCTION}
\medskip

The matter content of extragalactic systems, {\it e.g.,\/} galaxies
and clusters of galaxies, can be assessed in two independent ways: from the
dynamics of test objects, {\it e.g.,\/} extended rotation curves of disk
galaxies delineated by neutral hydrogen, or the velocity dispersion profile of
cluster delineated by galaxies, and from the deflection and focusing of
electromagnetic radiation, {\it e.g.,\/} gravitationally lensing cluster of
galaxies.  Because the ``bending of light'' by gravitational fields is
intrinsically a relativistic effect, the second approach provides a way to test
the relativistic aspects of gravitation at the extragalactic level. This is
particularly important in light of a variety of proposals (Tohline 1982,
Milgrom 1983a,b,c, Bekenstein \& Milgrom 1984, Sanders 1984, Bekenstein 1987,
Kuhn \& Kruglyak 1987, Sanders 1986a,b, 1988, 1989, Manneheim \& Kazanas 1989,
Kazanas \& Manneheim 1991, Sanders 1991, Bekenstein 1992, Romatka 1992, Milgrom
1993, Kenmoku et al. 1993) that seek to replace the dark matter paradigm with
one or another deviation of gravity from standard theory at extragalactic
scales.  Whatever the success of such schemes, and there have been many in the
case of Milgrom's, they must still face the challenge of providing a theory of
light bending that correctly predicts the features of gravitational lenses,
arcs, {\it etc.\/}

Now in the standard gravitational theory (general relativity [GR] and its
Newtonian limit), the two approaches to measure the matter content of a system
use the same tool to describe the predictions: the Newtonian gravitational
field $\vec g$.  Thus the acceleration of a test object in a galaxy or in a
cluster of galaxies is just this $\vec g$ evaluated at the position of the
object. And the bending angle of a light ray as it passes near the system along
an approximately straight path may be stated to lowest order in
$\upsilon^2/c^2$, where $\upsilon$ denotes the typical velocity in the system,
as  $$
\vartheta={2\over c^2}\int |g^\perp| dz   \eqno(e1)
$$
where $\perp$ denotes the component normal to the ray's direction, and
$dz$ is the element of length along the ray.  In this context  do
observations of cosmic gravitational lenses affirm the dark matter paradigm in
the context of standard gravity theory and rule out novel gravitational
theories as alternatives to dark matter ?

Unfortunately, up to now both observational and theoretical ambiguities
have stood in the way of a sharp answer to this question.  As detailed below,
the observations of lensing systems, principally clusters of galaxies,
do not yet permit a very accurate comparison of the lensing mass with
the standard dynamical mass derived from kinematical measurements.
And unconventional gravity theories have not, up to now, led to model
independent predictions for the lensing that can be confronted with the
evidence and the predictions from the dark matter picture.  For instance, it
has not been known in any generality whether in scalar--tensor (ST) type
gravitational theories without dark matter, the scalar field is expected to
bend light beyond what is accomplished by the luminous matter.  We now review
the status of the observations and unconventional gravity theoretical
frameworks.

Regarding observations of gravitational lenses, in  those cases of
strong imaging events {\it i.e.\/}, the formation of multiple images or
Einstein rings, where the lens is an individual galaxy (see Blandford \&
Naryan 1992), there is no indication of a discrepancy between the lensing and
luminous masses (Breimer \& Sanders 1993). This is not surprising because,
typically, the impact parameter of photons forming multiple images is only a
few Kpc, and  there is convincing evidence that the bright inner regions of
galaxies are typically dominated by the visible component (Kent 1987, van
Albada \& Sancisi 1987).  True, Maoz and Rix (1993) have claimed that the
observed frequency and image separation
of multiply-lensed quasars can only be achieved if galaxies
-- in particular elliptical galaxies -- have dark halos {\it i.e.\/}, the dark
component  makes a contribution to the image magnification and
separation.  However, this result
depends upon rather uncertain model parameters and scaling relations for
elliptical galaxies.

The issue of lensing by dark haloes has also been examined by observations of
 the lensing effects of forground galaxies on background galaxies within
projected separations between 10 kpc and 200 kpc {\it i.e.\/}, weak
gravitational imaging (Tyson et al. 1984). In this situation of large
projected separations, the presence of the foreground mass does not create
multiple images of a background galaxies, but does systematically distort
their shapes. This is very relevant to the
problem of the mass discrepancy because it is a light--bending probe of the
mass distribution in the outer parts of galaxies where the dark halo is thought
to dominate.  In the present context it is interesting that the results of
Tyson et al. are consistent with less extensive dark haloes than implied
by measures of the extended rotation curves.  For example, these results
imply that a typical lens galaxy has a total mass less than $3\times 10^{11}
\,M_\odot$ within a radius of 80 kpc; this is comparable to the luminous mass
of a typical elliptical galaxy  i.e. a galaxy with the fiducial
luminosity, $L^*$, in the Schechter (1976) luminosity function.
Is this result a falsification of
dark matter hypothesis ?  Unfortunately the Tyson et al. result remains
controversial. For example, Kovner \& Milgrom (1987) argue that the assumption
of infinite distance for the background sources underestimates the total mass
of the forground galaxies by a factor of three.  On the other hand, Breimer
(1993), using a realistic distribution of redshifts for the background sources,
has confirmed the result of Tyson et al. provided that the intrinsic source
size is larger than the seeing disk. Clearly, further observations of this
sort under conditions of good seeing would be most useful.  But it can be said
that, up to now, there is no conclusive evidence for lensing by dark
matter in individual galaxies.

The most persuasive evidence in favor of lensing by dark matter is provided by
 the luminous arcs seen in the central regions of a dozen rich clusters
(Blandford \& Naryan 1992).  It is now well-established that these are
background galaxies imaged into partial Einstein rings by the forground cluster
(Lynds \& Petrosian 1986, Soucail et al. 1987). The presence of one of these
arcs implies, via \Eq{e1}, the  existence of more than $10^{13}\, M_\odot$
projected within the inner 200 kpc (the typical Einstein ring radius), masses
which exceed, by a factor of ten or twenty, the projected luminous mass
(Bergmann, Petrosian \& Lynds 1990).  Moreover, the observations of
systematically distorted images of background galaxies  within projected radii
of 1/2 to one Mpc in two distant clusters (Tyson, Valdez, \& Wenk,
1990) {\it apparently} requires a substantial extended dark component. In the
cluster A 370, the most well--studied of the arc systems, the cluster velocity
dispersion predicted by the observed arc is consistent within the observational
errors with the observed velocity dispersion of the cluster galaxies, assuming
that the visible galaxies trace the dark mass distribution and that the
velocity distribution is isotropic (Mellier et al. 1988, Breimer \& Sanders
1992).  There thus appears to be consistency between the virial mass and the
lensing mass.  This would seem to be evidence in favor of the dark matter
hypothesis.

However, these observations do not yet rule out generic unconventional
gravity theories without dark matter for two reasons. First of all,
determinations of the mass and mass distribution by standard dynamical methods
are notoriously crude.  Virial methods, including those based upon the Jeans
equation, which use measurements of the line-of-sight velocities of cluster
galaxies are beset by uncertainties resulting from limited statistics, the
unknown degree of anisotropy in the velocity distribution, the the effects of
subclustering and contamination, and the unknown distribution of the total
(observed and unobserved) mass (The and White 1986, Merritt 1987).  The
potentially powerful method of using the observed density and temperature
distributions of the hot X-ray emitting gas has been limited by the lack of
high resolution spectral information; {\it i.e.,\/} the temperature
distribution is very uncertain (Sarazin 1988). For these reasons, dynamical
masses are typically determined to an accuracy of a factor of two at best.
Secondly, recent reanalyses of the gaseous component of clusters on the basis
of results from ROSAT and earlier X-ray observatories indicate that the
contribution of the hot gas to the total cluster mass may be greater than
previously supposed -- up to 40\% in some cases (Hughes 1989, Eayles et al.
1991, Briel, Henry \& Bohringer 1992, Bohringer, Schwarz \& Briel 1993).  This
goes a long way towards relieving the discrepancy in clusters.  Moreover, many
rich clusters show evidence for cooling flows which can deposit up to
$10^{13}\, M_\odot$ in the central regions in a Hubble time, presumably in the
form of cold gas or low mass stars (Sarazin 1988). This raises the possibility
that, in clusters containing giant arcs, the lensing mass is nothing more
exotic than hot gas or objects formed out of the hot gas. Future X-ray
satellites, which will probe the temperature distribution of the gas in
clusters, should clarify the issue of the mass distribution in the central
regions and the contribution of the hot gas to the total mass in the lensing
clusters.

In light of the above discussion, claims sometimes heard, {\it e.g.\/}
Dar (1993), that the extant observations of gravitational lenses, when compared
with dynamical determinations of the virial mass, certify the validity of
general relativity in the extragalactic regime are evidently premature. They do
not  reckon with the intrinsically crude determination of $\vec g$ provided by
the dynamical measurements.  Neither do they  take cognizance of the dearth, in
the extant literature, of clearcut predictions for gravitational light bending
in theories of gravity, which aim at explaining the data without an appeal to
dark matter.  There is a good reason for this dearth.  Bending of light in a
gravitational theory depends on details of the theory: what fields constitute
gravity, how are they coupled to gravitating matter, what equations are
satisfied by these fields, \dots ?  On the face of it, predictions of light
bending would have to be worked out separately for each theory, and it
certainly could not be ruled out {\it a priori\/} that some model theory could
produce light bending of the same strength as GR.

The actual situation is not so bad.  In many known gravity theories where a
scalar field plays a gravitational role, it does so by entering as a conformal
factor in the relation between the primitive (Einstein) metric and the
physical one.  This fact implies, in a rough way, that the scalar field will
not induce any light bending of itself (Bekenstein 1992, Romatka 1992). Two
advances were required in order to turn this observation into the solid
conclusion (this paper) that ST unconventional gravity theories
cannot reconcile the lensing and dynamical  (velocity dispersions) data for
clusters without requiring much ordinary or exotic unobserved matter.

One advance was to establish the generic way to couple a scalar field to
the metric.  Bekenstein (1993) has shown that the relation between Einstein
and physical metric may  be generalized to the form of a {\it disformal\/}
transformation in which the stretch of length in the special spacetime
direction delineated by the scalar field gradient is either larger or smaller
than the average one.  Because of the anisotropic stretch, it is no longer
automatic that the scalar field cannot cause bending of light.  The disformal
transformation, discussed in Sec.~2., seems to be the generic way to
introduce a scalar field into a gravity theory which has a fighting chance to
pass the tests of GR.  Considerations of causality dictate (Bekenstein 1992)
that the special direction undergoes a smaller stretch than the average over
all directions.  This particular sign turns out to be critical in the results
to be described presently.

The second advance required was a relativistic calculation of light
bending for a generic ST gravitational theory.  The rough insight that, as a
conformal factor in the  physical metric, a scalar field cannot bend light,
does not reckon with the fact that the stress energy of the scalar field can,
by way of its contribution to the sources of the Einstein metric, contribute
to bending of light.  This contribution cannot be ignored: the evidence is
that there is  a significant non-luminous mass in clusters, which could very
well turn out to be energy of coherent fields.  A truly relativistic
calculation of light bending will take into account this effect.  Such is
carried out in this paper, in Secs.~3. and 4. for the bending due to a
spherically symmetric system in a generic gravitational theory involving
scalar fields with positive energy densities.

The result (Sec.~5.) is surprising: despite all the subtleties mentioned,
the bending angle cannot exceed that predicted by the standard theory for the
actual matter (excluding the scalar field)  present in the system.  Thus  it
is a theorem that in a generic ST theory of gravity, the scalar field cannot
enhance lensing.  However, if a theory is presented as a candidate for
gravity with no dark matter, it will, almost by construction, predict that
the visible matter in the system generates a stronger gravitational field
than would be predicted by standard theory.  Therefore, modified
gravitational theories based on metric and scalar fields and assuming no dark
matter, and standard theory with dark matter must give distinctly
different predictions for the comparison of dynamical  and light bending
data.  That is, a modified gravity theory of the type we have in mind cannot
mimic all predictions of standard theory with dark matter.

Thus, as we conclude in Sec.6., if future measurements of lensing and of
velocity dispersions in clusters (or of the temperature distribution of the
hot gas) certify the presently popular view that gravitational lensing
requires the same mass discrepancies in galaxy clusters as are implied by
test body dynamical measurements, then unobserved matter makes up a large
fraction of the mass in such clusters, whether gravity is best described by
conventional theory or by some unconventional ST theory.  In the later case
the indication would be that in the cluster regime the predictions of the
unconventional theory differ little from those of conventional gravity
theory, with the difference between the theories being significant, if at
all, only on scale of galaxies.  By contrast, if future observations show
that the lensing by clusters suggest masses below those implied by the
velocity dispersions of galaxies or the distribution of hot gas, then dark
matter would loose much of its appeal as an explanation for the entire mass
discrepancy, while unconventional gravity theories of ST type would become
quite relevant. In this light gravitational lensing is seen to be a crucial
tool for the issue of dark matter vs. unconventional gravity.

In what follows we choose units such that the speed of light $c=1$, but  we
shall retain Newton's constant explicitly.  Our conventions for gravitational
theory are those of Misner, Thorne and Wheeler (1973). In particular we assume
metric signature $(-,+,+,+)$.

\bigskip
\centerline{ 2. CHARACTERIZING GRAVITATIONAL THEORIES}
\medskip
\centerline{2.1. {\it From General Relativity to Scalar--Tensor Theory\/}}
\medskip

Since we shall discuss gravitational theories which involve a scalar field, it
is in order to start with a discussion of why a scalar field is the most
natural entity that can be used to modify standard gravitational theory.  The
full characterization of a gravitational theory must be relativistic.  The
simplest such theory, GR, is formulated in terms of the metric
of spacetime, $g_{\alpha\beta}$, which enters into all dynamical equations for
matter in gravitational fields in the same manner that Minkowski's metric
enters in gravity's absence.  The simplicity of GR is
manifested in the absence of other gravitational fields apart from the metric,
and in the fact that the dynamical action for the metric is the simplest
invariant action that can be built in 4-D curved spacetime:   $$  S_{\rm
g}={1\over 16\pi G}\int{R\sqrt{-g}}d^4x  \eqno(e2)  $$ Here $R$ is the scalar
curvature built from the metric $g_{\alpha\beta}$ and its first and second
derivatives, and $g$ is the metric's determinant.

The success of GR in confronting the results of solar
system precision experiments (Will 1992) suggests that the correct
gravitational theory, if not identical to GR, must nevertheless
share with it some key features.  Therefore, any theory which attempts to
displace dark matter as an explanation of the mass discrepancy in large
astrophysical systems must show clear kinship with GR.  The
action (2) and the underlying metric $g_{\alpha\beta}$ are such key features of
GR that it is tempting to see them as required of any competing
theory.  We shall assume that a theory of gravity, to be credible, must have
action (2) in some choice of local units (in some conformal frame).

Similarly, the role of a curved spacetime metric in rewriting
special relativistic physics to include the effects of gravity seems a required
feature of theories that are to compete succesfully with GR.  We shall assume,
therefore, that the generic theory to be considered uses some such metric.
However, it does not follow that this last metric, call it $\tilde
g_{\alpha\beta}$, is identical to $g_{\alpha\beta}$.  In GR that is the case,
and the equivalence is a facet of the strong equivalence principle which is
part and parcel of GR.  However, it is only the Einstein equivalence principle,
a  weaker principle, which is supported by many experiments. As stressed by
Dicke long ago (Dicke 1962), those experiments are evidence only for the
existence of the metric $\tilde g_{\alpha\beta}$, and it is logically
consistent with  all extant experiments for the metrics $\tilde
g_{\alpha\beta}$ and $g_{\alpha\beta}$ to be distinct.  The metric
$g_{\alpha\beta}$, often referred to as the Einstein metric,  shall here be
called the {\it gravitational metric\/}; $\tilde g_{\alpha\beta}$ shall here be
referred to as the {\it physical metric\/}.  This last is the metric determined
by rods and clocks built of matter.

In the ancient Nordstr\"om theory of gravity (Nordstr\"om 1913), which actually
preceded GR, the metric $g_{\alpha\beta}$ was taken as flat,
whereas $\tilde g_{\alpha\beta}$ was taken as conformal to it, with the
conformal factor being the square of a scalar field which obeyed the standard
massless scalar equation written with $g_{\alpha\beta}$.  Thus Nordstr\"om's
theory is a pure scalar gravity theory. Despite its inherent simplicity,
the theory was rejected early.  It is amusing, given our concerns here, that
its failing was predicting that there is no bending of light.

The Bergmann--Wagoner scalar tensor gravitational theories (Will 1992), of
which Brans--Dicke  theory is the simplest, are a natural marriage of
Nordstr\"om's theory with GR.   The gravitational metric
$g_{\alpha\beta}$ is taken to satisfy Einstein--like equations, whereas the
physical metric $\tilde g_{\alpha\beta}$  is taken to be conformally related to
it, with the conformal factor being intimately related to the dynamical scalar
field of the theory, which together with the metric constitutes gravity.

\medskip
\centerline{2.2. {\it Disformal Transformation\/}}
\medskip

Of course, once a scalar field, $\psi$, is admitted into gravitational
theory, the relation between physical and gravitational metrics may be more
general.  For example, consider a disformal relation (Bekenstein 1992) between
the metrics,
$$
\tilde g_{\alpha\beta}=\exp(2\psi)\,[A(I) g_{\alpha\beta} + L^2
B(I)\psi,_\alpha\psi,_\beta]  \eqno(e3)
$$
where $L$ is a scale of length, and
$A$ and $B$ are two functions of the invariant
$$
I\equiv L^2\tilde
g^{\alpha\beta}\psi,_\alpha\psi,_\beta  \eqno(e4)
$$
The form of the factor $\exp(2\psi)$ is dictated by the reasonable requirement
that a shift in the zero of $\psi$ shall have no physical consequences. Indeed,
adding a constant to $\psi$ merely changes the global units of length paced out
by the physical metric.

The relation \(e3) is the most general between $\tilde g_{\alpha\beta}$ and
$g_{\alpha\beta}$ based on a scalar field $\psi$ which involves only the first
derivatives of $\psi$.  Generalizations of the relation  (3) involving second
derivatives of $\psi$ obviously entail a higher derivative theory, {\it i.e.\/}
one in which the matter field equations involve field derivatives of third or
higher order.  It is well known that  such theories tend to display causality
problems such as preacceleration and runaway solutions, and it seems wise to
exclude them at the outset. It has also been shown (Bekenstein 1993) that the
disformal relation \(e3) is the most general relation between a given
gravitational metric $g_{\alpha\beta}$ and a Finsler geometry for the matter
equations which is allowed by the requirements of causality and universality of
free fall.

Let us recapitulate the structure of the generic ST gravitational
theory.  The gravitational action (2) is responsible for dynamics of the
gravitational metric $g_{\alpha\beta}$.  The full Einstein--like equations are
obtained from the variation of the full action
$$
S_{\rm tot}=S_{\rm g}+S_{\rm m}+S_{\psi}  \eqno(e5)
$$
with respect to $g_{\alpha\beta}$. Here $S_{\rm m}$ is the action of
the matter fields obtained by replacing the Minkowski metric in the special
relativistic action for matter by $\tilde g_{\alpha\beta}$, and partial
derivatives by covariant derivatives constructed with $\tilde
g_{\alpha\beta}$.  And $S_\psi$ is an action for $\psi$ built out of $\psi$,
its first derivatives, and $g_{\alpha\beta}$.  The dynamics of $\psi$ comes
from its field equation which is obtained by varying $S_{\rm tot}$ with
respect to $\psi$.  They combine with the dynamics of $g_{\alpha\beta}$
coming from the Einstein--like equations, to determine how $\tilde
g_{\alpha\beta}$ varies in spacetime.

\medskip
\centerline{2.3. {\it Examples and Generic Theory\/}}
\medskip

Let us look at some examples of theories encompassed by the stated framework.
First, GR has $S_\psi=0$ with $\psi=0$ by fiat, with
$A=B=1$.  Here the two metrics are equivalent.  Brans--Dicke theory has $A=1$,
$B=0$ and
$$
S_\psi=-{1\over 16\pi G}\int (\omega+3/2)
{g^{\alpha\beta}\psi,_\alpha\psi,_\beta}\sqrt{-g} d^4x \eqno(e6)
$$
with  $\Lambda=e^\psi$ being the scalar field in Dicke's (nonmetric) form of
the theory (Dicke 1962), and $\omega$ being the usual  Brans--Dicke parameter
(really an inverse coupling constant in modern jargon).  As is well known,
Brans--Dicke theory is in good accord with all solar system tests of
gravitational theory for $\omega>10^3$.  The Bergmann-Wagoner theories differ
from the special Brans--Dicke case in that $\omega=\omega(\psi)$ and a
potential term is added to the kinetic part of the action \(e6).  In the
extragalactic context all the above theories are esentially indistinguishable.

Several ST theories have been proposed as alternatives to the dark
matter hypothesis coupled to standard gravity theory.  First in simplicity is
Sanders' two scalar field theory (Sanders 1986), which implicitly assumes $A=1$
and $B=0$, and in which a linear massless scalar field is accompanied by a
massive scalar field whose kinetic action has a sign opposite to the usual
one.  This negative energy field is repulsive.  The theory may pass the various
tests of GR, and is not obviously in contradiction with the
universality of free fall.  Yet, the negative energy is a detractive feature.

Following Sander's theory in complexity is the aquadratic lagrangian (AQUAL)
theory (Bekenstein and Milgrom 1984).  It has $A=1$, $B=0$ and
$$
S_\psi=-{1\over 8\pi GL^2}\int F(I)\sqrt{-g}d^4x  \eqno(e7)
$$
with $F(I)\propto I^{3/2}$ for small $I$ and $F\propto I$ for large $I$.  This
theory agrees with all solar system tests of relativity provided $F(I)$
appraches its linear form rapidly enough as $I$ rises.  It also reproduces much
of the mass discrepancy phenomenology of spiral galaxies as encapsulated in
Milgrom's MOND formula (Milgrom 1983a,b).  Sanders (1986) has proposed a
variant of AQUAL in which $F(I)$ also turns linear in $I$ for sufficiently
low $I$.

One problem with AQUAL and its variants is that the scalar field can
propagate superluminally (Bekenstein 1987, 1988, 1992).  This acausality is
eliminated in the phase coupled gravity (PCG) theory (Bekenstein 1987, 1988,
1992).  It has $A=1$ and $B=0$ and instead of one scalar field, it has two, the
usual $\psi$ and a second one ${\cal A}$.  The PCG dynamics comes from the
action  $$  S_{\psi,{\cal A}}=-{1\over 2}\int g^{\alpha\beta}[{\cal
A},_\alpha{\cal A},_\beta+\eta^{-2}{\cal A}^2\psi,_\alpha\psi,_\beta + V({\cal
A}^2)]\sqrt{-g}d^4x    \eqno(e7a)
$$
where $\eta$ is a small coupling constant and $V$ is a potential.  For small
$\eta$ the term in $S_{\psi,{\cal A}}$ involving $\psi,_\alpha$ dominates the
one involving ${\cal A},_\alpha$.  Neglecting the latter term eliminates ${\cal
A}$ as a dynamical variable.  Indeed, extremization of the action with respect
to ${\cal A}$ establishes that ${\cal A}$ is a function of
$g^{\alpha\beta}\psi,_\alpha\psi,_\beta$, so that to lowest order in $\eta$,
the PCG action reduces to AQUAL's, \Eq{e7}.  This suggests that PCG should be
as succesful as AQUAL in explaining the phenomenology of the mass
discrepancy.  Several in--depth studies (Bekenstein 1987, Sanders 1988,
1989) have shown that this is the case, but have also uncovered potential
problems with PCG in the solar system (Bekenstein 1987), and in galaxies
(Sanders 1988a).

{}From all the above examples we may write down the generic scalar action of
interest:
$$
S_{\psi,{\cal A}}=-\int {\cal E}({\cal I}, {\cal J}, {\cal K},
{\cal A})\sqrt{-g}d^4x \eqno(e46)
$$
Here ${\cal E}$ is a function, and
${\cal I}\equiv g^{\alpha\beta}\psi,_\alpha\psi,_\beta$,
${\cal J}\equiv g^{\alpha\beta}{\cal A},_\alpha{\cal A},_\beta$ and
${\cal K}\equiv g^{\alpha\beta}{\cal A},_\alpha \phi,_\beta$ are the
three invariants that can be formed from first derivatives of ${\cal A}$ and
$\phi$. Although it would not change any of the discussion to come, $\psi$ is
not included as an argument of ${\cal E}$   because it would spoil the
invariance of the theory under a shift of the zero of $\psi$.    At little
cost we could generalize our action to include any number of ${\cal A}$--type
fields.  Note that we do not assume in what follows that the kinetic part of
the scalar's action can be separated out, and is a quadratic form.  We
shall, however, assume that the function
${\cal E}({\cal I}, {\cal J}, {\cal K}, {\cal A})$ is such that
the scalar field always bears positive energy density. In Sec.~5 we
translate this requirement into conditions on ${\cal E}$ and its
derivatives.
 \bigskip
\centerline{ 3. LIGHT BENDING IN GENERAL}
\medskip
\centerline{3.1. {\it Equation for Light Rays\/}}
\medskip
In all modified gravity theories proposed so far as alternatives to dark
matter, $B=0$ and the two metrics are taken as conformally related.  Because
Maxwell's equations are conformally invariant, this seems to say that the
scalar field, because it enters only as a aconformal factor, has no influence
on the propagation of light (Bekenstein 1992, Romatka 1992).  Since the
gravitational metric comes from Einstein--like equations, this would seem to
say that all the ST theories with $B=0$ will give the same bending
of light as GR.  However, as mentioned in  Sec.~1.,
observations of arcs suggest that the lensing in clusters of galaxies is much
stronger than can be ascribed, via GR and formula \(e1), to the
visible matter.  This could mean that the modified gravity theories fail, and
there is much dark matter in clusters.  Alternatively, one can hope
(Bekenstein 1992) that theories with $B\not= 0$ could play a role in
reconciling the large lensing with the idea of unconventional gravity by
breaking the conformal relation between the metrics.  Below we shall see that
this hope is dashed because causality requires the ``wrong sign'' for $B$.

The argument from conformality of the metrics to the absence of a light
bending ascribable to the scalar field (Bekenstein 1992) leaves out an
important factor.  The scalar field, by virtue of its energy--momentum tensor,
must make a contribution to gravitational metric $g_{\alpha\beta}$, and thus
must lead to some extra light bending.  The calculations to follow are designed
to take this extra bending into account.  To accomplish this they are carried
out at the fully relativistic level, rather than starting from results like
formula \(e1) which are already stated in nonrelativistic terms.  The results
are surprising in that if the scalar field bears strictly positive energy, the
extra light bending is negative, {\it i.e.\/}, if anything it decrements the
general relativistic bending angle. As a result, no scalar--based gravity
theory can make exactly the same predictions as GR for both
dynamics of galaxies and clusters, and light bending.  The opportunity to
discriminate sharply between the two approaches thus arises.

We shall discuss ``light bending'' of either electromagnetic or neutrino
radiation.  The Maxwell, massless Dirac and Weyl equations are all
conformally invariant.  Hence the physics of light bending may be
discussed equally well by using the physical metric $\tilde g_{\alpha\beta}$
or the reduced metric
$$
\bar g_{\alpha\beta}=g_{\alpha\beta}+\varpi\psi,_\alpha\psi,_\beta
\eqno(e8)
$$
where $\varpi\equiv L^2B(I)/A(I)$.  It is important in all that
follows that elementary considerations of signature and causality
require that $A(I)>$ while $B(I)\leq 0$ (Bekenstein 1992, 1993).  Hence we may
take $\varpi\leq 0$.

For astrophysical systems the dimensions of the lensing system are large
compared to the wavelengths of interest.  Hence, we are only interested
in geometric optics results.  This means that our central question
concerns the form of  a {\it null\/} geodesic in the spacetime defined by the
reduced metric $\bar g_{\alpha\beta}$ (we assume that ``masses'' like the
neutrino mass or the plasma frequency in intergalactic space are small).
Therefore, the tracks of light rays through a gravitating cluster,
$x^\mu=x^\mu(\xi)$, where $\xi$ is a parameter along the ray, must satisfy
$$
\bar g_{\alpha\beta}{dx^\alpha\over d\xi}{dx^\beta\over d\xi}=0  \eqno(e9)
$$
together with any restrictions arising from symetries and conservation
laws.

\medskip
\centerline{3.2. {\it Bending in Spherical Symmetry\/}}
\medskip

It is sufficient for the points to be made to restrict  attention to light
bending by a static spherically symmetric configuration.  As is well
known, on symmetry grounds the metric of the corresponding spacetime,
say $g_{\alpha\beta}$, may always be put in the form
$$
ds^2=g_{\alpha\beta}dx^\alpha dx^\beta=-e^\nu dt^2+e^\lambda dr^2
+r^2(d\theta^2+\sin\theta^2 d\varphi^2) \eqno(e10)
$$
with $\nu$ and $\lambda$ functions of $r$ only. Similarly, we expect
$\psi=\psi(r)$ only.  Hence \Eq{e9} takes the form
$$
-e^\nu \left({dt\over d\xi}\right)^2+(e^\lambda+\varpi
\psi'^2)\left({dr\over d\xi}\right)^2 +r^2\left({d\theta\over
d\xi}\right)=0  \eqno(e11)
$$
where a prime denotes derivative with respect to $r$, and we have assumed
that by angular momentum conservation, the path is confined to the equatorial
plane.

Conservation of energy (static configuration) tells us that (Misner, Thorne
and Wheeler 1973)
$$
e^\nu{dt\over d\xi} = E  \eqno(e12)
$$
where $E$ is the constant ``energy'' of the ``photon''.  We write ``energy''
because $E$'s value depends on the scaling of the parameter $\xi$, which is
arbitrary. Likewise, conservation of angular momentum (spherical symmetry)
leads to
$$
r^2{d\theta\over d\xi} = \ell   \eqno(e13)
$$
with $\ell$ the ``photon'''s ``angular momentum''.  $\ell$ is likewise changed
by redefinition of $\xi$.  However, the ratio $b\equiv \ell/E$ is
independent of such redefinition, and may be interpreted as the impact
parameter of the ray from the center of the configuration.

Substituting \Eqs{e12,e13}  into \Eq{e11} and replacing $\xi$ by $\theta$ as
independent variable we cast the equation as
$$
\left({dr\over d\theta}\right)^2{\ell^2\over
r^4}(e^\lambda+\varpi\psi'^2)=e^{-\nu}E^2-{\ell^2\over r^2} \eqno(e14)
$$
Its first quadrature is
$$
\theta=\pm
\ell\int{{(e^\lambda+\varpi\psi'^2)^{1/2}\over(e^{-\nu}E^2-\ell^2/r^2)^
{1/2}}{dr\over r^2}}  \eqno(e15)
$$
This is the full description of the path of the ray in the form
$\theta=\theta(r)$.  The integral depends only on the ratio $b=\ell/E$, but not
separately on $E$ and $\ell$, thus realizing the independence on the parameter
$\xi$.

How is the bending angle $\vartheta$ determined ? Let $\theta=0$ stand for
the direction of incidence of the ray being bent so that the lower limit of the
integral in \Eq{e15} is $r=\infty$.  The integral from $r=\infty$ down to the
turning point value, $r=r_{\rm turn}$ where $d\theta/dr=0$, with minus sign
choice in the integral, is just half the rotation undergone by the radius
vector as the photon comes in from infinity to closest approach. Were there no
bending, the radius vector at closest approach would be at $\theta=\pi/2$.
Hence, when there is bending, the net bending angle is given by
$$
\vartheta=2\int_{r_{\rm turn}}^\infty{(e^\lambda+\varpi\psi'^2)^{1/2}
\over(e^{-\nu}(r/b)^2-1)^{1/2}}{dr\over r}-\pi  \eqno(e16)
$$  with the $\pi$
representing the overall change of $\theta$ when the path is perfectly
straight.

\medskip
\centerline{3.3. {\it Bending in Linear Approximation\/}}
\medskip

Equation~\(e16) is exact, but the integral cannot be evaluated analytically for
the  generic metric of interest.  However, some approximations that can make
the integration tractable are in order.  Recall that $\lambda$ and $\nu$ arise
from a solution $g_{\alpha\beta}$ of Einstein--like equations whose source is
the $T_{\alpha\beta}$ of the matter in the system as well as of the scalar
field generated by that matter.  Because the system in question is weakly
gravitating, we expect the metric $g_{\alpha\beta}$ to be nearly flat.  Hence
we may assume $|\lambda|, |\nu|\ll 1$.  Likewise, the physical metric $\tilde
g_{\alpha\beta}$ must be nearly flat by everyday experience in extragalactic
astronomy.  Of course, the flatness of the two metrics has to be evident in the
same set of coordinates.  Looking at \Eq{e3}, we see that the conditions for
common flatness of both metrics are that $|\varpi|\psi'^2\ll 1$ as well as
$A(I)\exp(2\psi)\approx 1$.  From $|\lambda|, |\nu|, |\varpi|\psi'^2\ll 1$ we
can obtain to linear order the following approximation to \Eq{e16}:
$$
\vartheta=2\int^\infty_{r_{\rm
turn}}{(1+\lambda/2+\varpi\psi'^2/2)\over
[(r^2/b^2)(1-\nu)-1]^{1/2}}{dr\over r}-\pi  \eqno(e17)
$$
It would, however, be premature to expand the integrand in $\nu$ because an
integral with $(r^2-b^2)^{3/2}$ in the denominator diverges at the turning
point.  Appendix A shows how to handle this subtlety, and how to compute the
integral to linear order in $\lambda$, $\nu$ and $\varpi$.  The result is
$$ \vartheta={b\over
2}\int^\infty_{-\infty}\left({\lambda+\varpi\psi'^2\over r^2}+{\nu'\over
r}\right)dz  \eqno(e17a)
$$
where an integration of the various quantities along
a straight path with impact parameter $b$ is meant.  Here $z$ stands for linear
distance.

\bigskip
\centerline{ 4. USING THE GRAVITATIONAL EQUATIONS}
\medskip
\centerline{4.1. {\it Correction to Standard Theory Bending\/}}
\medskip

We now determine $\lambda$, $\nu$ from the $T_{\alpha\beta}$ distribution
through the Einstein--like equations.  These are  $$
e^{-\lambda}(r^{-2}-r^{-1}\lambda')-r^{-2}=8\pi G T_t{}^t  \eqno(e27) $$ $$
e^{-\lambda}(r^{-1}\nu'+r^{-2})-r^{-2}=8\pi G T_r{}^r  \eqno(e28)
$$
Here $T_\alpha{}^\beta=g_{\alpha\gamma}T^{\gamma\beta}$, and $T^{\alpha\beta}$,
the
energy--momentum tensor,  is given by the variational derivative
$$
T^{\alpha\beta}\equiv -{2\over \sqrt{-g}}{\delta (S_{\rm m}+S_\psi)\over
\delta g_{\alpha\beta}}.  \eqno(e29)
$$
The solution of \Eq{e27} is
$$
e^{-\lambda}=1-2Gm(r)/r  \eqno(e30)
$$
with
$$
m(r)=-4\pi\int_0^r T_t{}^t r^2dr  \eqno(e31)
$$

Usually $m(r)$ is interpreted as the gravitating mass up to radius $r$.  Here,
since $T_t{}^t$  is defined with respect to the gravitational metric and not
to the physical one, $m(r)$ is mass in unconventional local units which  do
not correspond to those  measured by physical metersticks.  Appendix
B. develops the relation between the components of $T_\alpha{}^\beta$
in the two frames of units.  We find there that the factor that must
be included in the integrand of \Eq{e31} to give the physical mass is
very nearly unity for the situations of interest, so that $m(r)$ is very
nearly the physical mass.

If we now expand $e^{-\lambda}$ in \Eq{e28} to O$(\lambda)$, cancel a
$r^{-2}$ term, and neglect $\lambda \nu'/r$ in comparison with $\nu'/r$,
we get
$$
\nu'=\lambda/r+8\pi G r T_r{}^r  \eqno(e32)
$$
Substituting this and the linear approximation $\lambda=2Gm(r)/r$ in (26) gives
$$
\vartheta=2b\int_{-\infty}^\infty{G m(r)\over r^3}dz
+b\int_{-\infty}^\infty\left({\varpi\psi'^2\over 2r^2}+4\pi
GT_r{}^r\right)dz  \eqno(e33)
$$
Note that $Gm(r)/r^2$ is just the magnitude of the formal
Newtonian gravitational field $\vec g$ produced by all the mass energy interior
to the radius $r$.   The factor $b/r$ is just the cosine of the angle between
the radial direction and that normal to the ray at its nearest approach to the
center.  Hence the first integral in \Eq{e33} is identical to the standard
prediction for $\vartheta$, \Eq{e1} (except for the fact that the scalar
field's energy has to be included here in calculating $\vec g$). The second
integral gives the correction due to the scalar field which is not associated
with its energy.

We next recall that $\varpi\leq 0$ by causality.  Thus, even without
knowing anything about the equation that determines $\psi$, we can state
categorically that
$$
\vartheta\leq 2\int^\infty_{-\infty} |g^\perp| dz  + 4\pi
Gb\int^\infty_{-\infty}T_r{}^r\,dz\eqno(e34)
$$
We now proceed to break the integral over the stress $T_r{}^r$ into a
contribution $\delta\vartheta$ from  the matter (stars, gas, dark) and another,
$\Delta\vartheta$, from the scalar field, and to show that the first is always
negligible for the systems we have in mind.

\medskip
\centerline{4.2. {\it Evaluating the Stress Term\/}}
\medskip

The usual laws of energy--momentum conservation in special relativity must
here take the covariant form
$$
T_\mu{}^\nu{}_{;\nu}=0  \eqno(e35)
$$
where the semicolon signifies covariant differentiation with respect to
the gravitational metric $g_{\alpha\beta}$.   The law
\(e35) follows from coordinate invariance of the action $S_{\rm m}+S_\psi$
(Landau and Lifshitz 1975), or from the Bianch identities combined with the
gravitational field equations derived by varying $S_{\rm g}+S_{\rm
m}+S_\psi$ with respect to $g_{\alpha\beta}$. Now the $r$ component of
\Eq{e35} may be written in the form (Landau and Lifshitz 1975)
$$
(\sqrt{ - g}  T_r{}^r)'-{1\over 2}\sqrt{- g}(\partial g_{\alpha\beta}/\partial
r)  T^{\alpha\beta}=0      \eqno(e36)
$$
The assumed stationary and spherical symmetry forces $ g_{\alpha\beta}$ and
$ T^{\alpha\beta}$ to be symmetric tensors.  Thus \Eq{e36} takes the
form
$$
(e^{\lambda+\nu\over 2}r^2
T_r{}^r)'-{1\over 2}e^{\lambda+\nu\over 2}r^2\left[\nu' T_t{}^t+\lambda'
T_r{}^r + 2( T_\theta{}^\theta +  T_\phi{}^\phi)/r\right]=0 \eqno(e38)
$$
Spherical symmetry also demands $ T_\theta{}^\theta =  T_\phi{}^\phi$.
Substituting this and noting that the terms involving $\lambda'$ cancel
each other, we obtain the important result
$$
(e^{\nu/2}r^2
T_r{}^r)'={1\over 2}e^{\nu/2}r^2\left[\nu' T_t{}^t + 4
T_\theta{}^\theta/r\right] \eqno(e39a)
$$
As the radial component of the energy--momentum conservation equations,
\Eq{e39a} is the exact statement of hydrostatic equilibrium.

Let us integrate this last equation over $r$ in the interval [0,r].
Assuming that $T_r{}^r$ is bounded as befits  a pressure, we see that the
boundary term at $r=0$ vanishes.  Hence
$$
T_r{}^r(r)= {e^{-\nu/2}\over 2r^2}\int_0^r{r^2 e^{\nu/2}(\nu' T_t{}^t + 4
T_\theta{}^\theta/r)}dr \eqno(e39b)
$$
In our weakly gravitating cluster we may replace the factors $e^{\nu/2}$ by
unity to get
$$
T_r{}^r(r)\approx  {1\over 2r^2}\int_0^r r^2 (\nu' T_t{}^t +
4 T_\theta{}^\theta/r)dr \eqno(e39c)
$$
 which is our main working equation.

\medskip
\centerline{4.3. {\it Matter and Scalar Contributions\/}}
\medskip

There are two distinct contributions to $T_r{}^r$ which enter ultimately into
the expression for light bending, \Eqs{e33,e34}.  One belongs to the ordinary
matter, and the other to the scalar field.  First the ordinary matter (gas,
stars, dark) contributes to $ T_t{}^t$ and $ T_\theta^\theta$.  Because
this matter has a velocity dispersion $\sim \upsilon\ll 1$, we expect its
contribution  to the stress $| T_\theta^\theta|$ to be of order
$\upsilon^2| T_t{}^t|$.  To assess the magnitude of the first term in the
integral in \Eq{e39c}, let us write down the physical line element that
follows from \Eq{e10} in accordance with the transformation, \Eq{e3}:
$$
d\tilde s^2=-e^{\nu+2\psi}dt^2+ e^{\lambda+2\psi}dr^2+r^2
e^{2\psi}(d\theta^2+\sin^2\theta d\varphi^2)  \eqno(newmetric)
$$
It is evident from this that in the nonrelativistic limit
$\phi\equiv \nu/2+\psi$ plays the role of physical gravitational potential.
The anomalous term $\psi$ here encapsulates the effects of unconventional
gravity.  Any theory of gravity whose task is to explain the mass discrepancy
in terms of anomalous gravity must be such that $\psi' > 0$ (anomalous
contributions strengthens the gravitational force).  We may conclude
from this that since $\upsilon^2/r \sim \phi'$, $0 < \nu' <\upsilon^2/r$.
Hence, according to \Eq{e39c}, the matter's contribution to $T_r{}^r(r)$ is
$k \upsilon^2\langle\rho\rangle_r$, where  $\rho$ denotes the matter's energy
density, $(-T_t{}^t)_{\rm m}$, $k$ is a number of order unity and either
sign, and the average used is defined by
$$
\langle{\cal O}\rangle_{\rm r}\equiv {2\over r^2}\int_0^r r{\cal O}\,dr
\eqno(e40a)
$$

The scalar field makes equal contributions to $ T_t{}^t$ and $
 T_\theta^\theta$ in \Eqs{e39b,e39c}.  For according to \Eqs{e46,e29} the
scalar's energy--momentum tensor is
$$
\tau_\alpha{}^\beta=-{\cal E}g_\alpha{}^\beta+2(\partial {\cal E}/\partial{\cal
I})\psi,_\alpha\psi,^\beta +2(\partial {\cal E}/\partial{\cal J}){\cal
A},_\alpha{\cal A},^\beta +(\partial {\cal E}/\partial{\cal K})({\cal
A},_\alpha \psi,^\beta + \psi,_\alpha {\cal A},^\beta) \eqno(e47)
$$
Because of the symmetries, neither $\psi$ nor ${\cal A}$ depend on $t$,
$\theta$ or $\varphi$. Therefore,
$$
\tau_t{}^t=\tau_\theta^\theta=\tau_\varphi^\varphi=-{\cal E}  \eqno(new)
$$
where it is understood that ${\cal E}$ is to be evaluated using the solution
to the field equations for $\psi$ and ${\cal A}$.   Our assumption (Sec.~2)
of positive scalar field energy density obviously requires that
$-\tau_{tt}>0$, which in view of \Eq{e10} implies that $\tau_t{}^t < 0$ and
${\cal E}> 0$, at least for a static situation.

Because $\tau_t{}^t$ appears in \Eq{e39c} multiplied by $\nu'<
\upsilon^2/r$, the first term in the integral in \Eq{e39c} is negligible
compared to the second.  Replacing in this last $\tau_\theta^\theta$ by
$\tau_t{}^t$, and recalling the definition of averaging, we see that the
scalar's contribution to $T_r{}^r(r)$ is $\langle \tau_t{}^t
\rangle_r$.  We may summarize all the above in the equation
$$
T_r{}^r(r)=k \upsilon^2\langle\rho\rangle_r +
\langle\tau_t{}^t\rangle_r  \eqno(e47b)
$$
where $|k|$ may vary from system to system, and even with $r$ in accordance
with the distribution of velocity and density, but should usually be of order
unity.

\bigskip
\centerline{ 5. CONSEQUENCES OF POSITIVE ENERGY}
\medskip
\centerline{5.1. {\it Scalar Field Decreases Bending}}
\medskip

To assess the size of the line integral over $\langle\rho\rangle_r$ which
enters into \Eq{e34} by virtue of \Eq{e47b} we shall appeal to the Poisson
equation for the Newtonian gravitational field $\vec g_{\rm m}$ generated by
the matter.  In cylindrical coordinates this is
$$
{dg^\perp_{\rm m}\over db}+{g^\perp_{\rm m}\over b} +{dg^{z}_{\rm m}\over
dz}= - 4\pi G\rho  \eqno(e42)
$$
where we assume spherical symmetry to drop derivatives with respect to
$\varphi$, and denote by $b$ the coordinate in the direction of $\perp$.
Multiplying this equation by $b$ and integrating over $z$ gives
$$
\left(b{d\over db}+1\right)\int_{-\infty}^\infty g^\perp_{\rm m} dz
 = - 4\pi Gb\int_{-\infty}^\infty \rho dz  \eqno(e43)
$$
where we have assumed that $g^{z}\rightarrow 0$ as $|z|\rightarrow
\infty$.
Hence, taking into account that $\langle\rho\rangle_b$ is of the same order as
$\rho(b)$, we see that the contribution of the matter to the last term in
\Eq{e34} is
$$
\delta\vartheta\sim - k\upsilon^2\left(b{d\over db}+1\right)
\int_{-\infty}^\infty g^\perp_{\rm m} dz  \eqno(e44)
$$
Now, whatever the $b$ dependence of the last integral, its
logarithmic derivative in \Eq{e44} is unlikely to differ from itself by more
than an order of magnitude.  Since $\upsilon$ in a cluster is typically $10^3$
km s$^{-1} \approx 3\times 10^{-3}c$, it is clear that $\delta\vartheta$
may be neglected as compared with the first contribution to $\vartheta$ in
\Eq{e34}.  (In a relativistic cluster, {\it e.g.\/}
cluster of neutron stars at galactic center, $\upsilon\sim 1$ and
$\delta\vartheta$ would no longer be negligible).

According to \Eqs{e34,e40a,e47b}, the scalar field's contribution to the
bending angle through the last term in \Eq{e34} is
$$
\Delta\vartheta= 8\pi Gb\int_{-\infty}^\infty {dz\over r^2}\int_0^r
\tau_t{}^t r dr  \eqno(e50)
$$
It is already clear from the positivity of scalar field energy
$-\tau_t{}^t$ that $\Delta\vartheta<0$. As we shall see, this negative
contribution dominates the positive one of the scalar field to
the first term in $\vartheta$ as given by \Eq{e34}. We may transform the
inner integral in \Eq{e50} by appealing to the Poisson equation for the
formal Newtonian potential $\Phi_{\rm s}$ produced by the scalar's energy
density.  In spherical coordinates this is
$$ {1\over r}{d^2 (r\Phi_{\rm
s})\over dr^2}= - 4\pi G\tau_t{}^t  \eqno(e51)
$$
Multiplying through by $r$, integrating from $r=0$ to $r$, and noting that
$d\Phi_{\rm s}/dr$ must be finite at $r=0$, we find
$$
{4\pi G\over r^2}\int_0^r \tau_t{}^t r dr =
- {d\Phi_{\rm s}(r)/dr\over r}
+{\Phi_{\rm s}(0) - \Phi_{\rm s}(r)\over r^2}  \eqno(e52)
$$
Of course, since $\Phi_{\rm s}$ comes from \Eq{e51},
$d\Phi_{\rm s}(r)/dr=Gm_{\rm s}(r)/r^2$, where $m_{\rm s}(r)$ is the
scalar's contribution to the total mass, {\it c.f.\/} \Eq{e31}.  Thus if we
substitute \Eq{e52} in \Eq{e50} and that into \Eq{e34}, and compare the last
one with \Eq{e33}, we see that the scalar's contribution to the first
integral in \Eq{e34} is exactly cancelled out.  We are thus left with
$$
\vartheta\leq 2\int^\infty_{-\infty} |g^\perp_{\rm m}| dz
+2b\int^\infty_{-\infty}{\Phi_{\rm s}(0) - \Phi_{\rm s}(r)\over r^2}dz
\eqno(e53)
$$
where $\vec g_{\rm m}$ is the formal Newtonian field of
the matter alone.  Now because the scalar's energy density $-\tau_t{}^t
\geq 0$, it follows in the usual way from the Poisson equation \(e51) that
$\Phi_{\rm s}(r)$ grows with $r$.  Thus the second integral in \Eq{e53} is
strictly negative.  Hence,
$$
\vartheta < 2\int^\infty_{-\infty} |g^\perp_{\rm m}| dz \eqno(e54)
$$
This important result tells us that the scalar fields of the theory fail to
augment the light bending ability of the matter, and may even weaken it.

\bigskip
\centerline{5.2. {\it Divergent Gravitational Lenses ?}}
\medskip

What happens in the hypothetical case when the scalar field energy becomes
the dominant part of the gravitational lense's mass ?  We continue to assume
that the configuration is nonrelativistic in the sense that $\lambda$,
$|\nu|$, {\it etc.\/} are small compared to unity. Then the same
arguments as above lead to the cancellation of  the scalar's contribution to
the first integral in \Eq{e34}.  However, now we may neglect the matter's
contributions $\delta\vartheta$ and $\int^\infty_{-\infty} |g^\perp_{\rm m}|
dz$ as compared to the last term in \Eq{e52}.  The result is   $$
 \vartheta\leq 2b\int^\infty_{-\infty}{\Phi_{\rm s}(0) - \Phi_{\rm s}(r)\over
r^2}dz  \eqno(e55)
$$
which makes it clear that the bending angle is negative.  Such a
scalar--energy dominated equilibrium configuration would thus behave as a
divergent lense when bending light !

Under what conditions do equilibrium configurations dominated by scalar
fields  exist ? We may examine the question in the extreme case that
matter is negligible without making the nonrelativistic approximation by
returning to \Eq{e39b}, and replacing $ T_\alpha{}^\beta\rightarrow
\tau_\alpha{}^\beta$.  By using \Eq{new} we may recast \Eq{e39b}
into the form
$$  \tau_r{}^r(r)={e^{-\nu/2}\over r^2}\int_0^r({r^2 e^{\nu/2})'
\tau_t{}^t}dr \eqno(e56)
$$
Now the factor $r^2 e^{\nu/2}$ should grow with $r$ because the potential
$\nu/2$ is expected to increase with $r$  (attractive gravitational force) ,
and because of the increasing factor $r^2$.  Even if over some range
$\nu$ were to decrease with $r$, it seems highly unlikely that
$e^{\nu/2}$ would decrease faster than $r^{-2}$.  Thus positive energy density
of the scalar field means here that at all radii $\tau_r{}^r < 0$.

We now multiply  \Eq{e56} by $r^2 e^{\nu/2}$ and
differentiate with respect to $r$.  The result may be put in the form
$$
(\tau_r{}^r)'=(r^2 e^{\nu/2})'e^{-\nu/2}r^{-2}(\tau_t{}^t-\tau_r{}^r)
\eqno(e57)
$$
However, according to \Eq{e47}
$$
\tau_t{}^t-\tau_r{}^r=-2e^{-\lambda}\left[(\partial {\cal E}/\partial{\cal
I})\psi,_r^2 +(\partial {\cal E}/\partial{\cal J}){\cal
A},_r^2 + (\partial {\cal E}/\partial{\cal K}){\cal
A},_r \psi,_r \right]
\eqno(p1)
$$
We now argue that positive energy density of the scalar field with respect
to any observer implies that the quantity in square brackets in \Eq{p1}
is positive.  If $U^\alpha$ denotes the timelike 4--velocity of an
observer, the scalar field will exhibit positive energy density to him if
$T_{\alpha\beta}U^\alpha U^\beta > 0$. Since $U^\alpha U_\beta =
-1$, substituting \Eq{e47} into this condition we obtain the condition
$$
{\cal E} + 2\left[(\partial {\cal E}/\partial{\cal
I})(\psi,_\alpha U^\alpha)^2 +(\partial {\cal E}/\partial{\cal J})({\cal
A},_\alpha U^\alpha)^2 + (\partial {\cal E}/\partial{\cal K}){\cal
A},_\alpha U^\alpha \psi,_\beta U^\beta \right] > 0
\eqno(p2)
$$
Now, we already know that ${\cal E} > 0$. Because $U^\alpha$
occurs in the quadratic form but not in ${\cal E}$, positivity of this
expression for any timelike $U^\alpha$ is guaranteed only if the quadratic
form is positive definite.  This requires
$$
\partial {\cal E}/\partial{\cal I} > 0;\qquad \partial {\cal
E}/\partial{\cal J}> 0 \eqno(p3)
$$
and
$$
(\partial {\cal E}/\partial{\cal K})^2 < 4(\partial {\cal E}/\partial{\cal
I}) (\partial {\cal E}/\partial{\cal J})
\eqno(p4)
$$

{}From these three conditions it follows that the quadratic form in the
square brackets in \Eq{p1} is positive for any $\psi,_r$ and ${\cal
A},_r$.  But then $\tau_t{}^t-\tau_r{}^r < 0$, so that it follows from
\Eq{e57} that not only is $\tau_r{}^r$ negative for all $r$, but it also
decreases outward.   In particular, $\tau_r{}^r(r) < -|\tau_r{}^r(0)|$ .
This, of course, means that $\tau_r{}^r$ can never reach zero, as would
befit a finite configuration.  This behavior  also means, by \Eq{p1}
that
$$
-\tau_t{}^t(r) >
|\tau_r{}^r(0)|+2e^{-\lambda}\left[(\partial
{\cal E}/\partial{\cal I})\psi,_r^2 +(\partial {\cal E}/\partial{\cal
J}){\cal A},_r^2 + (\partial {\cal E}/\partial{\cal K}){\cal A},_r \psi,_r
\right] > 0 \eqno(p5)
$$
which makes it clear that the energy density cannot vanish asymptotically as
would be required for a configuration of finite mass.  We conclude that a
static spherically symmetric self--gravitating configuration of scalar field
is untenable. It seems likely that the result would still hold for a
configuration containing matter in which the scalar field dominates the mass
of the matter.  This makes the existence of divergent gravitational lenses
very doubtful.

\medskip
\centerline{ 6. CONCLUSIONS AND CAVEATS}
\medskip

Comparing \Eqs{e1,e54} we conclude that in a gravitational theory where
gravity is mediated only by a metric and some  scalar fields, the bending of
light by a weakly gravitating system, like a galaxy or a cluster of galaxies,
cannot exceed the bending predicted by GR for the mass of
visible and hitherto undetected matter (but excluding the scalar's energy).
This means that {\it if gravity is properly described by a ST
theory, and one uses the standard theory's formula \(e1) to interpret
gravitational lensing observations, one can only  underestimate the mass
present in stars, gas and dark matter.\/}

The same conclusion obtains within GR if matter includes one or more scalar
fields, {\it i.e.\/} Higgs fields.  We obtain this case in our formalism by
setting $A(I)=1$, $B(I)=0$ and $\psi=0$.  Then $\tilde
g_{\alpha\beta}=g_{\alpha\beta}$.  The ${\cal A}$ fields may stand for the
scalar fields in question, and we may take over the results of our previous
discussion: {\it if gravity is described by GR, but matter includes various
scalar fields, use of formula \(e1) to interpret gravitational lensing
observations  can only underestimate the mass of matter present not in scalar
fields.\/}

Thus the observational claim that clusters of galaxies deflect light
more strongly than would be expected from the observable matter contained by
them, if it survives, cannot be interpreted in terms of scalar-tensor gravity
without dark matter. Specifically {\it if follow--up observations eventually
certify that the matter distribution inferred via standard theory from the
lensing is very much like that determined from the dynamics of test objects
or the temperature distribution of the X--ray emitting gas, then departures
from standard gravity of ST form cannot play a role in the
inner parts of clusters of galaxies which act as the lenses\/}.

A similar conclusion {\it cannot\/} be drawn for galactic gravitational lenses
since lensing by single galaxies, if  observed at all, is produced by the
inner regions where, in all likelyhood, the visible matter is all there is.
Neither is our last conclusion necessarily at variance with Milgrom's MOND
scheme since the accelerations in the cores of rich clusters seem  to be large
compared to $10^{-8}$ c.g.s., the scale at which anomalous gravity effects
set in according to MOND.  In fact, what we have demonstrated is that ST
unconventional gravity theories are irrelevant for understanding
gravitational lensing by galaxy clusters.  This does not exclude the
possibilities that this type of theory may be of relevance for understanding
the mass discrepancy in disk galaxies, or that the proper relativistic
formulation of ideas like MOND involves more esoteric gravitational physics,
{\it e.g.\/} modification of Newton's second law (Milgrom 1983a, 1993).

In fact, it must be stressed that the very fact that  a lensing mass, when
determined by standard theory,  exceeds the directly detectable mass in stars
or gas does not necessarily falsify unconventional gravity theories in
general.  There may be an as yet undetected component in clusters.  A
significant fraction, perhaps as large as 40\%, of the mass in clusters,
which was once considered to be dark, is now known to be in the form of hot
gas.  The matter in cooling flows must go somewhere, and such flows can
deposit a substantial mass  in the central regions of clusters. It would be
interesting if clusters containing arcs all showed evidence for cooling
flows.

A system in which the classical dynamical mass, determined from virial
methods or the distribution of hot gas, significantly exceeds the lensing
mass as determined by GR, would be very problematic for
the dark matter picture, but would be entirely consistent with
unconventional ST gravity.  This is not an entirely hypothetical
situation.  As mentioned in Sec.~1., the Tyson {et. al\/} measurements may
be an example of this at the galactic scale.   At cluster scales we may
point out that what has been noticed in the sky are the striking examples of
distant rich clusters containing luminous arcs or arclets;  of equal
importance may be the many examples of distant rich clusters {\it without\/}
arcs or arclets (too low lensing masses).

Our main results, {\it e.g.\/} \Eqs{e53,e54} were obtained under the
assumption that the scalar field configuration is truly static.  There is
thus a loophole in our conclusions.  An equilibrium configuration
containing a coherent complex scalar field oscillating harmonically in
time, as in the boson star models (Liddle and Madsen 1992), is not covered by
\Eq{e53}.  Neither is an equilibrium configuration in the PCG theory where
scalar energy is nonnegligible, and where the cosmological expansion causes
the field $\psi$ to vary in time in approximately linear fashion (Sanders
1989).  In future work we shall examine these situations.

\bigskip

We both are grateful for the hospitality of the Aspen Center for Physics where
this work was begun, and to Mordehai Milgrom for many conversations.  J.D.B.
thanks Rainer Romatka for comments and clarifications.

\endtitlepage
\vskip 0.5cm \centerline{ APPENDIX A.}
\medskip

Let us rewrite \Eq{e17} as
$$
\vartheta=-4\left[{\partial\over\partial\alpha}\int^\infty_{r_{\rm
turn}}(1+\lambda/2+\varpi\psi'^2/2)[(r^2/b^2)(1-\nu)-\alpha]^
{1/2}r^{-1}dr\right]_{\alpha=1}-\pi \eqno(e18)
$$
and then expand to first order in $\nu$.  The results are
$$
\vartheta=\vartheta_1+\vartheta_2  \eqno(e19)
$$
with
$$
\vartheta_1=\int^\infty_b{2+\lambda+\varpi\psi'^2
\over(r^2/b^2-1)^{1/2}}{dr\over r}-\pi  \eqno(e20)
$$
$$
\vartheta_2=2\left[{\partial\over
\partial\alpha}\int^\infty_{b\surd\alpha}{r\nu/b^2\
\over(r^2/b^2-\alpha)^{1/2}}\right]_{\alpha=1} \eqno(e21)
$$

We proceed to simplify the expression for $\vartheta_1$.  Two terms
cancel out because of the integral
$$
2\int^\infty_b{dr/r\over
(r^2/b^2-1)^{1/2}}=\pi  \eqno(e22)
$$
Next, we transform the independent variable from $r$ to
$z=\pm(r^2-b^2)^{1/2}$ which is just the Euclidean length along a ray
whose impact parameter is $b$, with zero length taken at the point of
closest approach to the center, $r=b$.  Thus
$$
\vartheta_1={b\over 2}\int^\infty_{-\infty}{\lambda
+\varpi\psi'^2\over b^2+z^2}dz  \eqno(e23)
$$
where we have extended the integral to $z=-\infty$ and divided by $2$.

We now turn to $\vartheta_2$.  Integration by parts converts it to
$$
\vartheta_2={2\over b}\left\{{\partial\over
\partial\alpha}\left[\lim_{r\rightarrow\infty}\nu(r^2-\alpha b^2)^{1/2}
-\int^\infty_{\surd\alpha b}\nu'(r^2-\alpha
b^2)^{1/2}dr\right]\right\}_{\alpha=1}\eqno(e24)
$$
By expanding the square root under the limit in $\alpha b^2/r^2$, we
verify that after differentiation with respect to $\alpha$, the
corresponding term vanishes.  Carrying out the $\alpha$ differentiation
of the integral, setting $\alpha=1$, and passing from variable $r$ to
$z$ we have
$$
\vartheta_2={b\over 2}\int^\infty_{-\infty}{\nu'\over
(b^2+z^2)^{1/2}}dz  \eqno(e25)
$$
Therefore, the full bending angle is
$$
\vartheta={b\over
2}\int^\infty_{-\infty}\left({\lambda+\varpi\psi'^2\over
r^2}+{\nu'\over r}\right)dz  \eqno(e26)
$$
where again $r\equiv (b^2+z^2)^{1/2}$.
\endtitlepage
\vskip 0.5cm
\centerline{ APPENDIX B.}
\medskip

We begin by writing the tilde analog of \Eq{e29}.
$$
\tilde T^{\alpha\beta}\equiv -{2\over \sqrt{-\tilde g}}{\delta (S_{\rm
m}+S_\psi)\over \delta \tilde g_{\alpha\beta}}.  \eqno(s29)
$$
In order to simplify the algebra we restrict ourselves to the case where $A$
and $B$ vary slowly, and neglect their derivatives.  Since according to \Eq{e3}
$$
\delta g_{\alpha\beta}/\delta
\tilde g_{\alpha\beta}= A^{-1}e^{-2\psi}  \eqno(s30)
$$
we find by the chain rule that
$$
\tilde T^{\alpha\beta}=(g/\tilde g)^{1/2}A^{-1}e^{-2\psi}T^{\alpha\beta}
\eqno(s31)
$$

Of greater interest are the mixed components  $\tilde T_\alpha{}^\beta$.
Contracting both sides of \Eq{s31} with $\tilde g_{\gamma\alpha}$ and using
the relation \(e3) and the definition $\varpi\equiv B/A$ we find that
$$
\tilde T_\gamma{}^\beta=(g/\tilde g)^{1/2}\left(T_\gamma{}^\beta
+ \varpi T^{\alpha\beta}\psi,_\alpha\psi,_\gamma\right) \eqno(s32)
$$
where indeces of $T^{\alpha\beta}$ are lowered with $g_{\alpha\beta}$ and
those of $\tilde T^{\alpha\beta}$ with $\tilde g_{\alpha\beta}$.

The above result is general.  For a static spherically symmetric situation
the two metrics are given by \Eqs{e10,newmetric} from which we infer that
$$
(g/\tilde g)^{1/2}=A^{-2}e^{-4\psi}(1+e^{-\lambda}\varpi\psi'^2)^{-1/2}
\eqno(s33)
$$
Further, because of the symmetries, $\psi,_\alpha=\psi,_r
\delta_\alpha{}^r$, and the mixed tensor $T_\gamma{}^\beta$ is diagonal.
Hence we find
$$
\tilde T_t{}^t/T_t{}^t=\tilde T_\theta{}^\theta/T_\theta{}^\theta = \tilde
T_\varphi{}^\varphi/T_\varphi{}^\varphi= A^{-2}e^{-4\psi}(1+e^{-\lambda}
\varpi\psi'^2)^{-1/2} \eqno(s34)
$$
and
$$
\tilde T_r{}^r/T_r{}^r= A^{-2}e^{-4\psi}(1+e^{-\lambda}
\varpi\psi'^2)^{1/2} \eqno(s35)
$$
Of course all the above results apply to the scalar field energy momentum
tensor $\tau_\alpha{}^\beta$ alone.

We may also use \Eq{s34} to reexpress \Eq{e31} for the mass function $m(r)$
in terms of the physical energy density $-\tilde T_t{}^t$:
$$
m(r)=-4\pi\int_0^r A^2 e^{4\psi} (1+e^{-\lambda}\varpi\psi'^2)^{1/2}
\tilde T_t{}^t r^2 dr  \eqno(s36)
$$
Since in the weak field situations under consideration $A e^{2\psi}\approx
1$, $\lambda\ll 1$ and $|\varpi|\psi'^2\ll 1$, it is clear that $m(r)$ is
very nearly the physical mass interior to radius $r$.
\endtitlepage
\parindent = -10pt

Bekenstein, J.D. and Milgrom, M. 1984, ApJ 286, 7.

Bekenstein, J. D. 1987, in Second Canadian Conference on General Relativity and
Relativistic Astrophysics, ed. A. Coley, C. Dyer, \& T. Tupper (Singapore:
World Scientific), 68.

Bekenstein, J.D. 1989, in  Developments in General Relativity,
Astrophysics and Quantum Theory, ed.  F.I. Cooperstock, L.P. Horwitz,
 \& J. Rosen, (Bristol: IOP Publishing), 156.

Bekenstein, J.D. 1992, in  Proceedings of the Sixth Marcel
Grossman Meeting on General Relativity, ed. H. Sato \& T. Nakamura
(Singapore: World Scientific), 905.

Bekenstein, J. D. 1993, Phys Rev D 48, 3641.

Bergmann, A.G., Petrosian, V. and Lynds, R. 1990, ApJ 350, 23.

Blandford, R. D. and Narayan, R. 1992, Ann Rev A\&A 30, 311.

Bohringer, H., Schwarz, R. A. and Briel, U. G. 1993, PASP.

Breimer, T. G. 1993, Dissertation, University of Groningen.

Breimer, T. G. and Sanders, R. H. 1992, MNRAS 257, 97.

Breimer, T. G. and Sanders, R. H. 1993, A\&A, 274, 96.

Briel, U. J., Henry, J. P. and Bohringer, H. 1992, A\&A 259, L31.

Dar, A. 1993, Technion preprint 1992.

Dicke, R. H. 1962, Phys Rev 125, 2163.

Eyles, C. J., Watt, M. P., Bertram, D., Church, M. J., Ponman, T. J.,
Skinner, G. K. and Willmore, A. P. 1991, ApJ 376, 23.

Hughes, J. P. 1989, ApJ 337, 21.

Kazanas D.and  Manneheim, P. D.  1991, ApJS,76, 431.

Kenmoku, M., Kitajima, E., Okamoto, Y., and Shigemoto, K. 1993, Int J Mod Phys
D 2, 123.

Kent, S.M. 1986, AJ 91, 1301.

Landau, L. D. and Lifshitz, E. M., 1975 Classical Theory of Fields
(London: Pergamon).

Lynds, R. and Petrosian, V. 1986, BAAS 18, 1014.

Liddle, A. R. and Madsen, M. S. 1992, Int J Mod Phys D, 1, 101.

Manneheim, P. D.  and Kazanas, D. 1989, ApJ 342, 635.

Maoz, D. and Rix, H-W. 1993, ApJ 416, 425.

Mellier, Y., Soucail, G. and Mathez, G. 1988, A\&A 199 13.

Merritt, D. 1987, ApJ 313, 121

Milgrom, M. 1983a, ApJ 270, 365.

Milgrom, M. 1983b, ApJ 270, 371.

Milgrom, M. 1983c, ApJ 270, 384.

Milgrom, M. 1993, Weizmann Institute preprint.

Misner, C. W. , Thorne K. S. and Wheeler J. A. 1973, Gravitation
(San Francisco: Freeman).

Nordstr\"om, G. 1913, Ann Phys (Leipzig) 42, 533.

Kent, S. M. 1987, AJ 93, 816.

Kovner, I. and Milgrom, M. 1987, ApJ 321, L113.

Kuhn, J. R.  and Kruglyak, L. 1987, ApJ 313, 1.

Romatka, R. 1992, Dissertation, Max Planck Institute
for Physics, Munich.

Sanders, R. H. 1984, AA 136, L21.

Sanders, R. H. 1986, AA 154, 135.

Sanders, R. H. 1986, MNRAS 223, 559.

Sanders, R. H. 1988, MNRAS 235, 105.

Sanders, R. H. 1989, MNRAS 241, 135.

Sanders, R. H. 1990, Ast Ap Rev 2, 1.

Sarazin, C. L. 1988, X-ray Emissions from Clusters of Galaxies
(Cambridge: Cambridge University).

Schechter, P. L. 1976, ApJ 203, 297.

Soucail, G., Mellier, Y., Fort, B., Hammer, F. and Mathez, G. 1987,
A\&A 172, L14.

The, L. S., and White, S. D. M. 1986, AJ 92, 1248.

Tyson, J. A., Valdez, F., Jarvis, J. F. and Mills, A. P. Jr. 1984,  Ap
J Letters, 281, L59.

Tyson, J. A., Valdez, F., and Wenk, R. A. 1990, ApJ 349, L1.

Tohline, J. E. 1982, in Internal Kinematics and
Dynamics of Galaxies, ed. A. Athanassoula (Dordrecht: Reidel), 205.

van Albada, T.S. and Sancisi, R. 1987, Phil Trans Roy Soc London
A 320, 447.

Weinberg, S. 1972, Gravitation and  Cosmology (New York: Wiley).

Will, C. M. 1992, Int J Mod Phys D 1, 113.

\end